\documentclass[xcolor=table]{article}
\usepackage{ijcai18}

\usepackage{times}
\usepackage{soul}
\usepackage[utf8]{inputenc}
\usepackage[small]{caption}

\usepackage{multirow}
\usepackage[table,xcdraw]{xcolor}

\usepackage{ucs}
\usepackage{amsmath}
\usepackage{amsfonts}
\usepackage{amssymb}
\usepackage{makeidx}
\usepackage{breqn}

\usepackage{url}
\usepackage{hyperref}
\usepackage{graphicx}
\usepackage{booktabs}
\usepackage{listings}
\usepackage{color}

\definecolor{dkgreen}{rgb}{0,0.6,0}
\definecolor{gray}{rgb}{0.5,0.5,0.5}
\definecolor{mauve}{rgb}{0.58,0,0.82}

\title{Machine Learning-based Variability Handling in IoT Agents}

\author{
Nathalia Nascimento$^1$$^2$, 
Paulo Alencar$^2$, 
Carlos Lucena$^1$$^2$ 
and Donald Cowan$^2$ 
\\ 
$^1$ Laboratory of Software Engineering. \\Pontifical Catholic University of Rio de Janeiro. Rio de Janeiro, Brazil.\\
$^2$ David R. Cheriton School of Computer Science. \\University of Waterloo. Waterloo, ON, Canada.\\
{nnascimento,lucena}@inf.puc-rio.br,
{palencar,dcowan}@csg.uwaterloo.ca
}

\begin{document}

\maketitle

\begin{abstract}
	Agent-based IoT applications have recently been proposed in several domains, such as health care, smart cities and agriculture. Deploying these applications in specific settings has been very challenging for many reasons including the complex static and dynamic variability of the physical devices such as sensors and actuators, the software application  behavior and the environment in which the application is embedded. In this paper, we propose a self-configurable IoT agent approach based on feedback-evaluative machine-learning. The approach involves: i) a variability model of IoT agents; ii) generation of sets of customized agents; iii) feedback evaluative machine learning; iv) modeling and composition of a group of IoT agents; and v) a feature-selection method based on manual and automatic feedback.

\end{abstract}
%

\section{Introduction}\label{sec:introduction}

Based on the Google Trends tool \cite{googletrends}, the Internet of Things (IoT) \cite{atzori2012social} is emerging as a topic  that is highly related to robotics and machine learning. In fact, the use of learning agents has been proposed as an appropriate approach to modeling IoT applications \cite{do2017fiot}. These types of applications address the problems of distributed control of devices that must work together to accomplish tasks \cite{atzori2012social}. This has caused agent-based IoT applications to be considered for several domains, such as health care, smart cities, and agriculture. For example, in a smart city, software agents can autonomously operate traffic lights \cite{do2017fiot,santos2017model}, driverless vehicles \cite{herrero2008decentralized} and street lights \cite{do2017engineering}.

Agents that can interact with other agents or the environment in which the applications are embedded are called {\itshape embodied agents} \cite{brooks1995intelligence,marocco2007emergence,Nolfi2016,do2017engineering}. The first step in creating an embodied agent is to design its interaction with an application's sensors and actuators, that is, the signals that the agent will send and receive \cite{Nolfi2016}. As a second step, the software
engineer provides this agent with a behavior specification compatible with its body and with the task to be accomplished. However, to specify completely the behaviors of a physical system at design-time and to identify and foster characteristics that lead to beneficial collective behavior is difficult \cite{mendoncca2017cooperative}. To mitigate these problems, many approaches \cite{marocco2007emergence,oliveira2014symbol,Nolfi2016,do2017engineering} have proposed the use of evolving neural networks \cite{nolfi1996learning} to enable an embodied agent to learn to adapt their behavior based on the dynamics of the  environment \cite{nolfi1996learning}.

The ability of a software system to be configured for different contexts and scenarios is called {\itshape variability} \cite{galster2014variability}. According to \cite{galster2014variability}, achieving variability in software systems requires software engineers to adopt suitable methods and tools for representing, managing and reasoning about change.

However, the number and complexity of variation points \cite{pohl2005software} that must be considered while modeling agents for IoT-based systems is quite high \cite{ayala2015software}. Thus, ``current and traditional agent development processes lack the necessary mechanisms to tackle specific management of components between different applications of the IoT, bearing in mind the inherent variability of these systems"
\cite{ayala2015software}.

 In this paper, we propose a self-configurable IoT agent approach based on feedback-evaluative machine-learning. The approach involves: (i) a variability model for IoT agents; (ii) generation of sets of customized agents; (iii) feedback-evaluative machine-learning; (iv) modeling and composition of a group of IoT agents; and (v) a feature-selection method based on both manual and automatic feedback.

\subsection{Motivation: Variability in IoT Agents}

\begin{table*}[!htb]
	\centering
	\caption{IoT Agents Variability.}
	\label{table:variabilities}
	\begin{tabular}{|c|c|l|l|}
		\hline
		\multicolumn{2}{|c|}{\cellcolor[HTML]{C0C0C0}}                                                                                                                                                                                                                            & \multicolumn{2}{c|}{\cellcolor[HTML]{C0C0C0}Behavior Variability}                                                                                                                                                                                                                                                                                                                                                                                                             \\ \cline{3-4} 
		\multicolumn{2}{|c|}{\multirow{-2}{*}{\cellcolor[HTML]{C0C0C0}Body Variability}}                                                                                                                                                                                          & \multicolumn{1}{c|}{\cellcolor[HTML]{EFEFEF}\begin{tabular}[c]{@{}c@{}}Behavior/\\ Constraint\\ Variability\end{tabular}}                                                                                                                                                                                  & \multicolumn{1}{c|}{\cellcolor[HTML]{EFEFEF}{\color[HTML]{000000} \begin{tabular}[c]{@{}c@{}}Analysis Architecture\\ (Neural Network Variability)\end{tabular}}} \\ \hline
		\multicolumn{2}{|c|}{Number of sensors}                                                                                                                                                                                                                                   & \multicolumn{1}{c|}{\begin{tabular}[c]{@{}c@{}}Number and type of \\ communication signals\end{tabular}}                                                                                                                                                                                                   & Number layers                                                                                                                                                    \\ \hline
		\multicolumn{2}{|c|}{\begin{tabular}[c]{@{}c@{}}Type of sensors (e.g. temperature,\\ humidity, motion, lighting, gases)\end{tabular}}                                                                                                                                     & \multicolumn{1}{c|}{\begin{tabular}[c]{@{}c@{}}Notification types \\ (e.g. alerts)\end{tabular}}                                                                                                                                                                                                           & \begin{tabular}[c]{@{}l@{}}Number neurons\\ per layer\end{tabular}                                                                                               \\ \hline
		\cellcolor[HTML]{EFEFEF}                                   & \cellcolor[HTML]{EFEFEF}\begin{tabular}[c]{@{}c@{}}Calibration of sensors \\ (e.g. temperature detector \\ range, range of \\ presence detection,\\ reaction time, range\\ of colors detection)\end{tabular} & \multicolumn{1}{c|}{\begin{tabular}[c]{@{}c@{}}Thresholds to activate\\ notifications\end{tabular}}                                                                                                                                                                                                        & \begin{tabular}[c]{@{}l@{}}Activation \\ Function (e.g.\\ linear, sigmoid)\end{tabular}                                                                         \\ \cline{2-4} 
		\cellcolor[HTML]{EFEFEF}                                   & \cellcolor[HTML]{EFEFEF}Energy Consumption                                                                                                                                                                   & {\color[HTML]{000000} }                                                                                                                                                                                                                                                                                    & \begin{tabular}[c]{@{}l@{}}Properties (e.g.\\ WTA, feedback)\end{tabular}                                                                                        \\ \cline{2-2} \cline{4-4} 
		\multirow{-3}{*}{\cellcolor[HTML]{EFEFEF}\textbf{Sensors}} & \cellcolor[HTML]{EFEFEF}Battery life                                                                                                                                                                         & {\color[HTML]{000000} }                                                                                                                                                                                                                                                                                    &                                                                                                                                                                  \\ \cline{1-2}
		\multicolumn{2}{|c|}{Communication device}                                                                                                                                                                                                                                & {\color[HTML]{000000} }                                                                                                                                                                                                                                                                                    &                                                                                                                                                                  \\ \cline{1-2}
		\multicolumn{2}{|c|}{\begin{tabular}[c]{@{}c@{}}Range of communication \\ devices (e.g. short range, \\ long range)\end{tabular}}                                                                                                                                         & {\color[HTML]{000000} }                                                                                                                                                                                                                                                                                    &                                                                                                                                                                  \\ \cline{1-2}
		\multicolumn{2}{|c|}{Number and type of motors}                                                                                                                                                                                                                           & {\color[HTML]{000000} }                                                                                                                                                                                                                                                                                    &                                                                                                                                                                  \\ \cline{1-2}
		\multicolumn{2}{|c|}{}                                                                                                                                                                                                                                                    & {\color[HTML]{000000} }                                                                                                                                                                                                                                                                                    &                                                                                                                                                                  \\
		\multicolumn{2}{|c|}{\multirow{-2}{*}{\begin{tabular}[c]{@{}c@{}}Number and type of \\ actuators (e.g. alarm)\end{tabular}}}                                                                                                                                              & \multirow{-7}{*}{{\color[HTML]{000000} \begin{tabular}[c]{@{}l@{}}IoT Application Logic - \\ connection between the\\ inputs and outputs (e.g. if the \\ lighting\_sensor value is zero,\\ then turn on the light, if the \\ temperature\_sensor is below \\ zero, then turn on the heater)\end{tabular}}} & \multirow{-6}{*}{\begin{tabular}[c]{@{}l@{}}Architecture (e.g.\\ full connected, output layer \\ connected to all of the hidden \\ units)\end{tabular}}          \\ \hline
	\end{tabular}
\end{table*}

In an Internet of Things application suite, there are several options for physical components and software behaviors for the design of a physical agent \cite{del2017piecewise,ayala2015software}. According to existing experiments \cite{vega2016beauty,soni2017smart} and our experience with the IoT domain \cite{do2015modeling,briot2016multi,do2016iot,do2017fiot,do2017engineering}, we introduce possible variants  of an IoT embodied agent in Table \ref{table:variabilities}. For example, the physical devices may vary in terms of the types of sensors, such as temperature and humidity, and in terms of actuators. Each sensor can also vary in terms of brands, changing such parameters as  energy consumption and battery life. The three main variation points we have identified as shown in Table \ref{table:variabilities} illustrate  the complexity of IoT agent-based applications.

 Thus, the complexity of the behavior of the agent will vary based on the physical components that are operated by  the agent. For example, if an agent is able to activate an alarm, which kinds of alerts can this agent generate? If this agent is able to communicate, how many words is this agent able to communicate? If this agent is able to control the temperature of a room, what are the threshold values set to change the room’s temperature?

In addition, we also need to deal with variants in agent architecture that the agent uses to sense the environment and  behave accordingly. For example, this architecture can be a decision tree, a state machine or a neural network. Many approaches \cite{marocco2007emergence,Nolfi2016,do2017engineering} use {\itshape neuroevolution}, which is ``a learning algorithm which uses genetic 
algorithms to train neural networks" \cite{whiteson2005evolving}). This type of network determines the behavior of an agent automatically based on its physical characteristics and the environment being monitored. A neural network is a well-known approach to provide responses dynamically and automatically, and create a mapping of input-output relations \cite{haykin1994neural}, which may compactly represent a set of ``if..then" conditions \cite{do2017engineering}, such as: ``if the temperature is below 10$^{\circ}$C, then turn on the heat." However, finding an appropriate neural network architecture based on the physical features and constraint behavior that were selected for an agent, is not easy. To model the neural network, we also need to account for its architectural variability, such as the activation function, the number of layers and neurons and properties such as the use of winner-take-all (WTA) as a neural selection mechanisms \cite{fukai1997simple} and the inclusion of recurrent connections \cite{marocco2007emergence}.

With respect to variabilities, \cite{marocco2007emergence}, performed two experiments with the same embodied agents, varying only the neural network architectures and neural activation functions. In the first experiment, they used a neural network without internal neurons, while in the second experiment, they used a neural network with internal neurons and recurrent connections. In addition, they also used different functions to compute the neurons' outputs. Based only on the neural network characteristics, they classified the robots from the first experiment as reactive robots (i.e. ``motor actions can only be determined on the basis of the current sensory state"), and non-reactive robots (i.e. ``motor actions are also influenced by previous sensory and internal states"). \cite{marocco2007emergence} analyzed whether the type of neural architecture influenced the performance of a team of robots. They showed that the differences in performance between reactive and non-reactive robots vary according to the environmental conditions and how the robots have been evaluated.

\cite{oliveira2014symbol} investigated symbol representations in communication based on the neural architecture topology that is used to control an embodied agent. They found that the communication system varies according to how the hidden layers connect the visual inputs to the auditory inputs.

These findings have helped us to conclude that to support the design of IoT embodied agents, we need to account for the variability of the physical body, the behavior constraints, and the architecture that analyses the inputs.

\section{Approach} \label{sec:approach}
\begin{figure}[!htb]
	\centering
	\includegraphics[width=8.2cm]{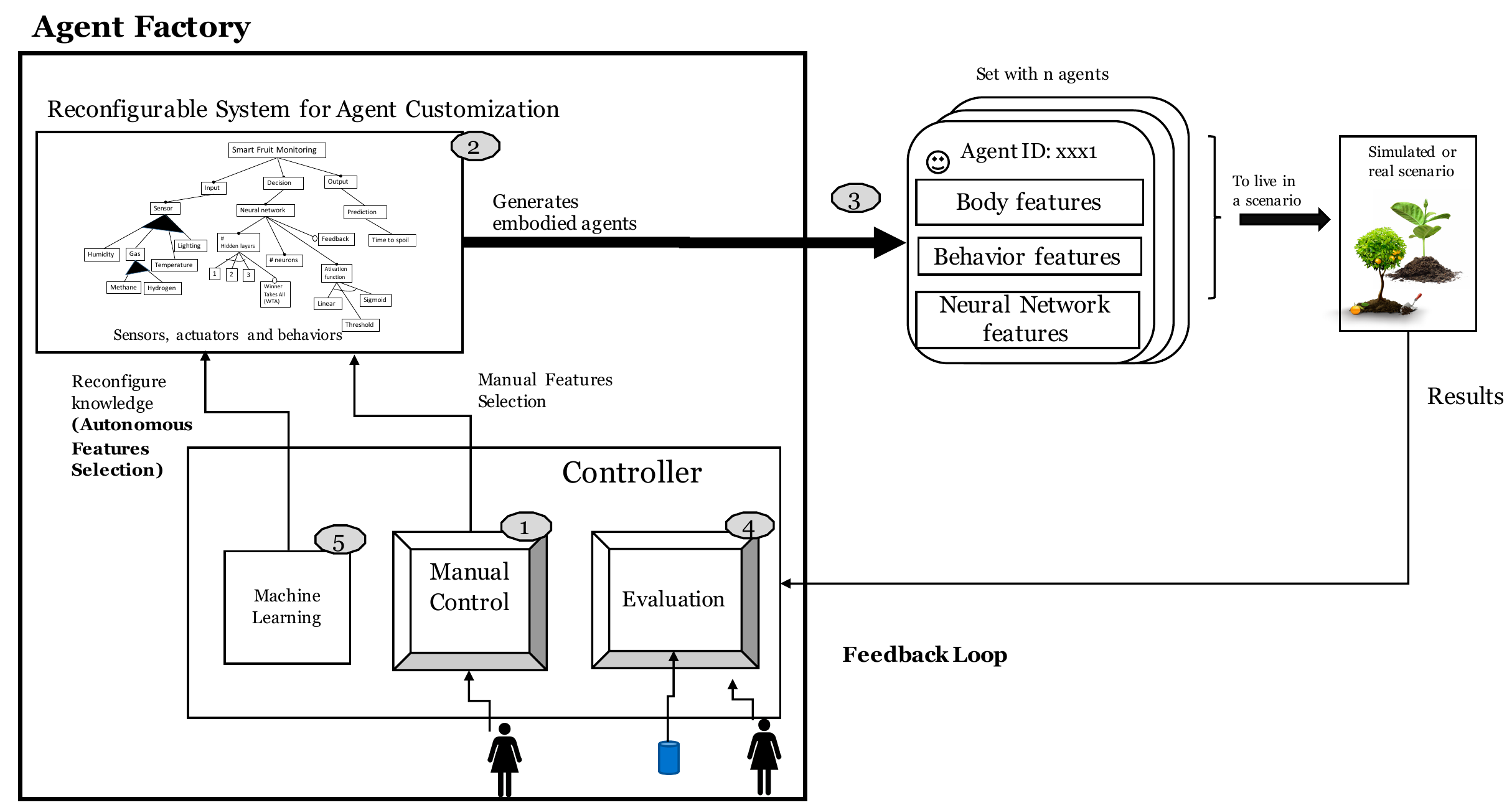}
	\caption{High-level model of the self-configurable agent approach to generate embodied agents. }
	\label{figure:approach}
\end{figure}

We aim to support the development of IoT embodied agents by designing a platform to support i) handling variability in IoT embodied agents, ii) selecting the physical components that will compose each agent, and iii) finding their appropriate behavior according to their bodies and the scenario where they will be applied. Figure \ref{figure:approach} depicts the high-level model of our proposed approach to self-configurable agents.

Basically, this platform or agent factory contains five modules: i) a manual control that allows an IoT expert to select the first set of features manually; ii) a reconfigurable system that contains the features that can be used to compose the set of agents incorporating feature-oriented domain analysis (FODA) \cite{pohl2005software} to model the software's variability; iii) the creation of a set of agents containing the selected features that are also able to use a neural network to learn about the environment; iv) a module for evaluating feedback tasks, by investigating the performance of the group of agents in the application scenario after the learning execution (depending on the evaluation result, the control module can trigger the machine learning algorithm to reconfigure the set of features); and v) a machine-learning module to select autonomously a new set of physical, behavior and neural network features.

\subsection{Current Implementation}

The current implementation of our architecture consists of two main parts. First (subsection 2.1.1), a human-in-the-loop selects the set of physical, behavior and neural network features for the group of IoT agents. Second (subsection 2.1.2), based on the features that were selected by the human, a neuroevolution-based algorithm is used to remove the irrelevant physical features and discover the agent's behavior.

During the second step, the neuroevolution-based algorithm considers a specific environment to discover the appropriate behaviors that enable a set of agents to achieve a collective task on that environment. After finding an appropriate behavior (i.e. the weights and topology of the neural network), the initial phase of the learning process is complete.

 However, an unexpected change in environment may force this process to be re-executed as all variation points can be affected. If this environmental change makes it necessary to add a new sensor to the agents' body, the way that the agents perceive the environment may also be reconfigured, and the learning process in the second step will also need to be re-executed. In addition, if the environment changes dynamically, there is a need to identify which variation points will be affected and how to handle the associated variability.

\subsubsection{2.1.1 Changing / Adding Features - Changing the search space for the neuroevolution-based algorithm} \label{sec:step1}

According to the FODA notation, features can be classified as mandatory, optional and alternative. Alternative features are not to be used in the same instance, such as the range of communication devices, the number of words to be communicated or the maximum number of hidden layers. For example, in the beginning of the experiment, if we select a neural network as the decision architecture, we must choose one of the features that defines the maximum number of hidden layers that this neural network can have, such as ``two" or ``three." So, if ``two" is selected, the search space for learning will be limited to the use of two hidden layers. If the communication system of the agents is limited to one word, the learning algorithm will not be able to test other solutions that could involve the communication of more than one word.

 Thus, the current search space to be used by the learning algorithm has been limited by the set of features that were selected to compose the embodied agents. However, there are three situations for which this search space may need to be changed or expanded: i) the learning algorithm does not find a good solution using this set of features, making it necessary to select alternative choices for some features (i.e. selecting a different activation function for the neural network) to reconfigure the set of agents; ii) the user changes some requirements of the agent-based system, making it necessary to add new unpredicted features to the feature model, as described in \cite{sharifloo2016learning}; and iii) the learning algorithm found an appropriate solution for the agents in the environment (i.e. the collection of agents are achieving their tasks in the application environment), but the environment changed dynamically, unexpectedly decreasing the performance of the agents.

In this step, there is a need to control the search space that will be used by the neuroevolution-based algorithm for the next step (described in subsection 2.1.2). This control consists of selecting the set of features to compose the system. For instance, a human-in-the-loop has performed this selection and reconfiguration. But our goal (and we designed our architecture for this purpose) is to enable an automatic reconfiguration of the system. In such a case, if the agents face an unexpected environmental change, a learning algorithm can be used to select a new set of features to compose the group of agents and execute the neuroevolution-based algorithm again. In this situation, we proposed the use of a learning algorithm to reconfigure a neural network (i.e. selecting another activation function), which can be seen as an automatic machine-learning approach (Auto-ML) \cite{muneesawang2002automatic}.

\subsubsection{2.1.2 Using neuroevolution to discard irrelevant features and discover the agent's current behavior} \label{sec:step2}

As described previously, each agent contains a neural network to make decisions. The weights, the topology, the input and output features of this neural network are determined based on an evolutionary algorithm. This algorithm makes changes based on the performance evaluation of the agents in the application.

\begin{figure}[!htb]
	\centering
	\includegraphics[width=6.1cm]{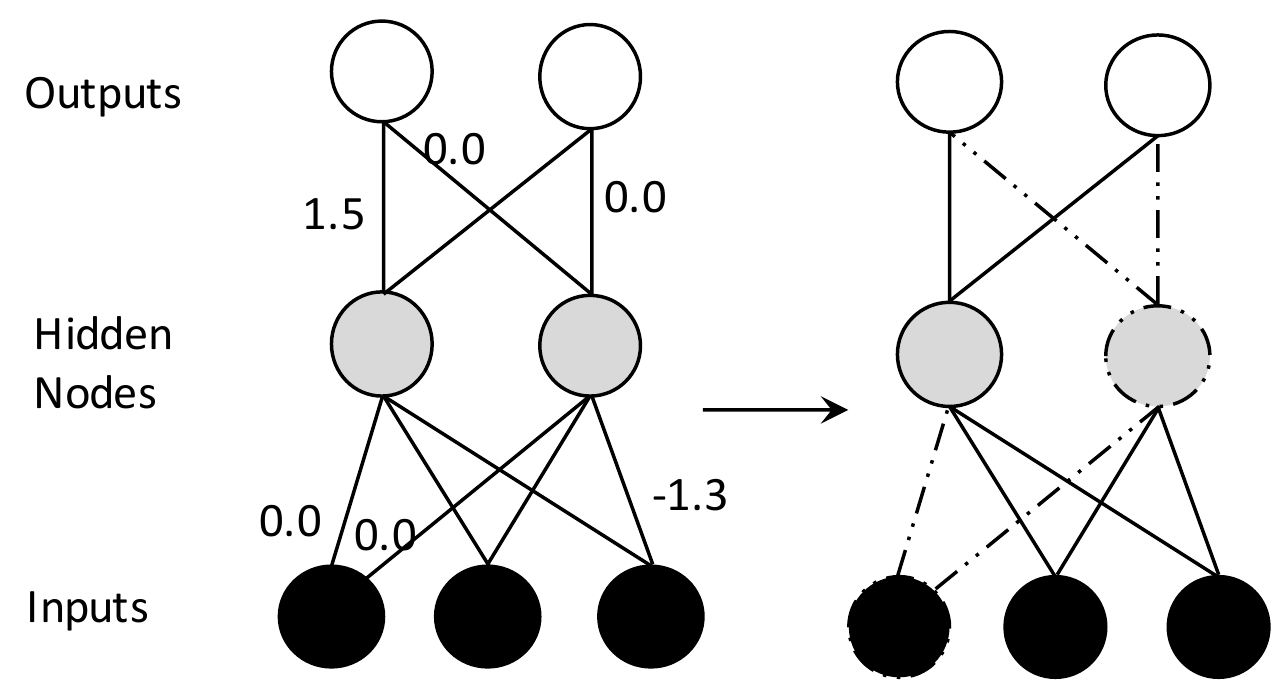}
	\caption{Removing input features and other neuronal connections. }
	\label{figure:neuroevolution}
\end{figure}

We implemented this neuroevolution algorithm based on the Feature Deselective NeuroEvolution of Augmenting Topologies (FD-NEAT) proposed by \cite{tan2009automated}. But instead of starting with a minimal architecture with inputs directly connected to the output layers, without a hidden layer as proposed by the traditional NEAT and FD-NEAT methods, we started with a three-layer neural network with all connections. In addition, we decided that a connection removal means a zeroed weight between two neurons, as illustrated in \ref{figure:neuroevolution}.

If the weight is zero, in our implementation, the neural network's output will not influence the activation function of the next neuron. In this case, the hidden layer will always exist, which can make the search more complicated. To mitigate this complexity, we established positive and negative thresholds for the weight setting in order to stimulate the connection removals. So, only the connections with higher contributions will remain during the evolutionary process. For example, in a weight range of [-2;+2], connections with ``0.2" or ``-0.1" weights are examples of connections that will be removed. In such a case, if all connections between a sensor input and the hidden layer are removed, this input feature will be discarded.

\section{Illustrative Example: Smart Street Lights}

To illustrate the variability dimensions of an IoT agent-based application, we selected and implemented one of the simplest examples from the IoT domain: a smart street light application. Even in a simple experiment of lighting control, we found many different prototypes in the literature \cite{carrillo2013lighting,de2016intelligent,do2017engineering}. For example, \cite{carrillo2013lighting} provided lights with cameras for image processing, while \cite{de2016intelligent} provided them with ambient light sensors, and \cite{do2017engineering} provided lights with ambient light and motion-detection sensors.

In this scenario, we consider a set of street lights distributed in a neighborhood. These street lights need to learn to save energy while maintaining the maximum visual comfort in the illuminated areas. For more details concerning this application scenario, see \cite{do2017engineering}.

\subsection{Feature-Oriented Domain Analysis (FODA)}

\begin{figure}[!htb]
	\centering
	\includegraphics[width=8.8cm]{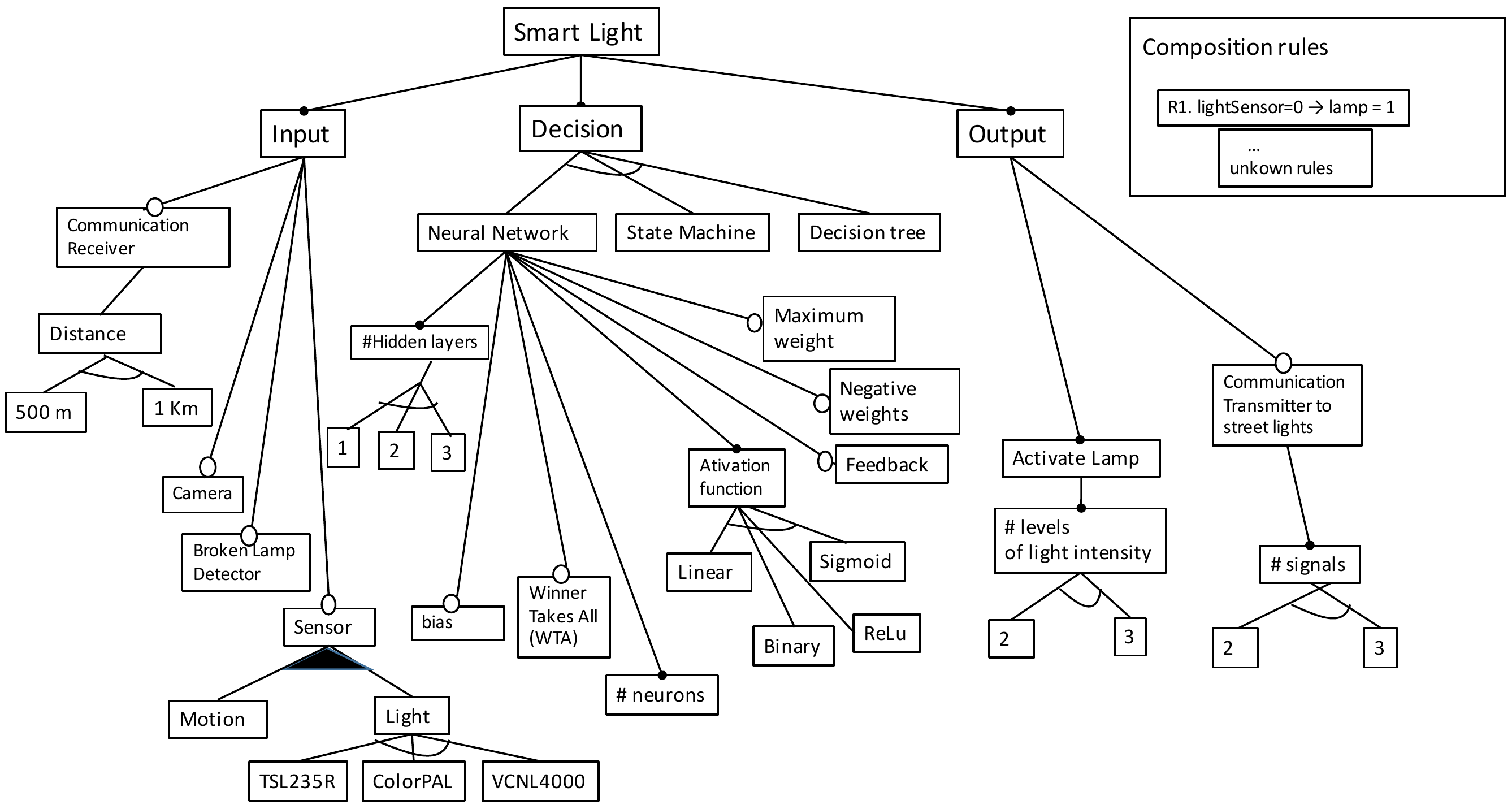}
	\caption{Feature model of a smart light agent. }
	\label{figure:neuralfeatures}
\end{figure}

Figure \ref{figure:neuralfeatures} illustrates the use of FODA to express IoT agent variability in a public lighting application.

As shown in this figure, even in a simple IoT agent, you may need to consider many variation points to create an IoT agent. According to the model, the input, decision and output are mandatory features. But the selection of sensors to compose the body of the agent is optional. If you decide to use sensors, you must select at least one of the sensors, such as the light sensor. In addition, if you select the light sensor feature, you must select which brand will be used. Depending on the selected light sensor brand, your agent will be able to sense very small changes in light or detect a full range of colors \cite{lightsensors}.

\subsection{Selecting Physical and Neural Network Features}

An IoT expert selected three physical inputs and two physical outputs to measure and operate each one of the street lights. The expert also added one behavior output: namely, the agents could ignore messages received from neighboring street lights. In addition, the engineer selected a neural network with one hidden layer with five units as the initial network for each agent with the sigmoid function as the activation function of this neural network.

\begin{figure}[!htb]
	\centering
	\includegraphics[width=5.1cm]{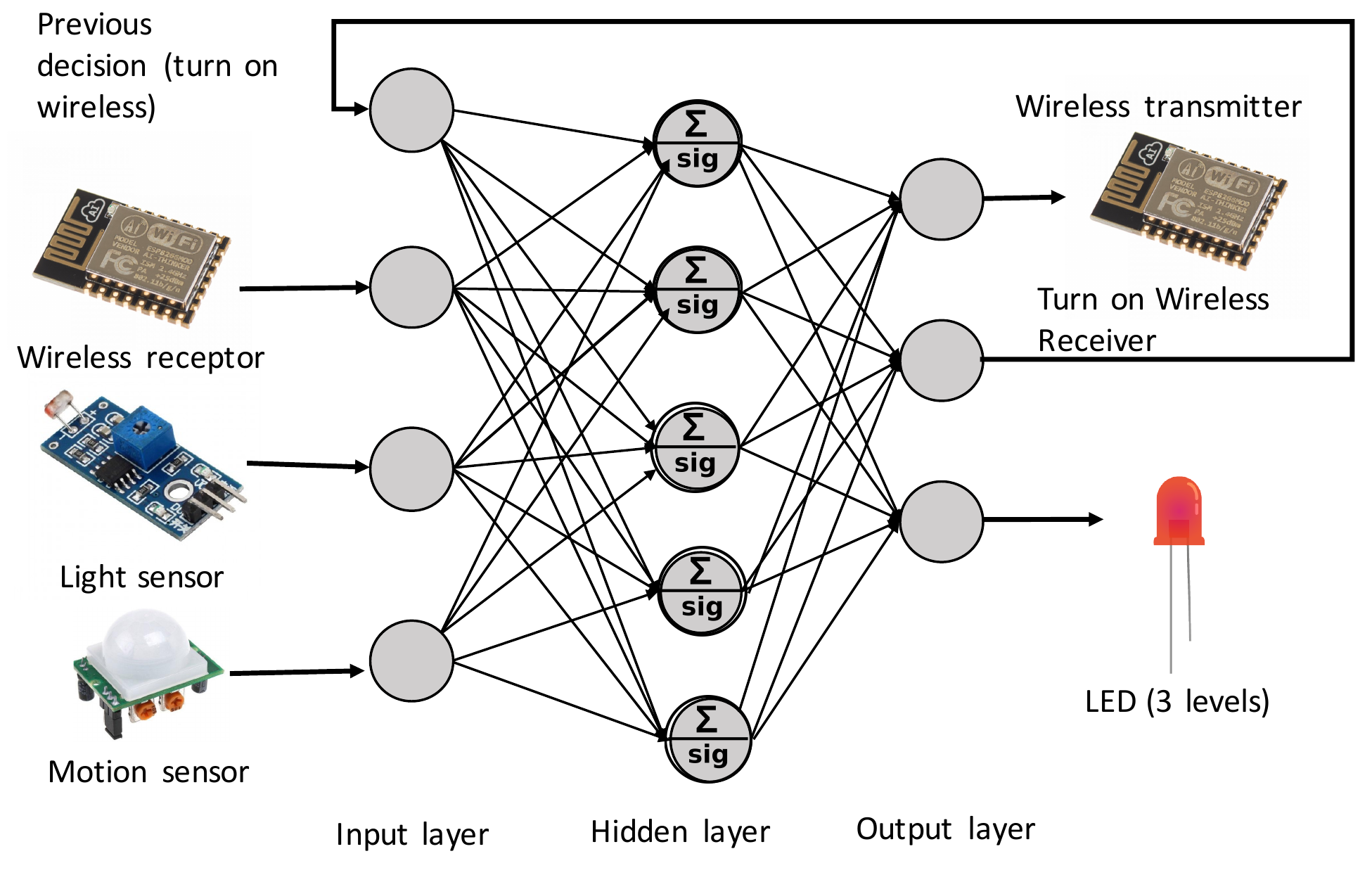}
	\caption{Neural network resulted from the first feature-selection interaction.}
	\label{figure:fruitneural}
\end{figure}

Figure \ref{figure:fruitneural} depicts the three-layer neural network that was generated based on the selected features. The input layer includes four units that encode the activation level of the sensors and the previous output value of the listening decision output. The output layer contains three output units: (i) listeningDecision, that enables the smart lamp to receive signals from neighboring street lights in the next cycle; (ii) wirelessTransmitter, a signal value to be transmitted to neighboring street lights; and (iii) lightDecision, that switches the light's OFF/DIM/ON functions.

\subsection{Learning about  the environment}
During the training process, the algorithm evaluates the options for weights of the network based on energy consumption, the number of people that finished their routes before the simulation ends, and the total time spent by people moving during their trip. Therefore, each weight-set trial is evaluated after the simulation ends based on the following equations:

\begin{equation}
pPeople = \frac{(completedPeople \times 100)}{totalPeople}
\label{eq:percentPeople}
\end{equation}
\begin{equation}
pEnergy = \frac{(totalEnergy \times 100)}{(\frac{11 \times (timeSimulation \times totalSmartLights)}{10})}	
\label{eq:percentEnergy}
\end{equation}
\begin{equation}
pTrip =\frac{(totalTimeTrip \times 100)}{((\frac{3 \times timeSimulation}{(2)}) \times totalPeople)}
\label{eq:percentTrip}
\end{equation}
\begin{equation} fitness = (1.0 \times pPeople) -\\ (0.6 \times pTrip) -\\ (0.4 \times pEnergy)
\label{eq:fitness}
\end{equation}

in which \begin{math} pPeople \end{math} is the percentage of people that completed their routes by the end of the simulation out of the total number of people participating in the simulation; 
\begin{math} pEnergy \end{math} is the percentage of energy that was consumed by street lights out of the maximum energy value that could be consumed during the simulation. We also considered the use of the wireless transmitter to calculate energy consumption; 
\begin{math} pTrip \end{math} is the percentage of the total duration time of people's trips out of the maximum time value that their trip could consume; and \begin{math} fitness \end{math} is the fitness of each representation candidate that encodes the neural network.

\subsubsection{Environmental Setting}
\begin{figure}[!htb]
	\centering
	\includegraphics[width=7.7cm]{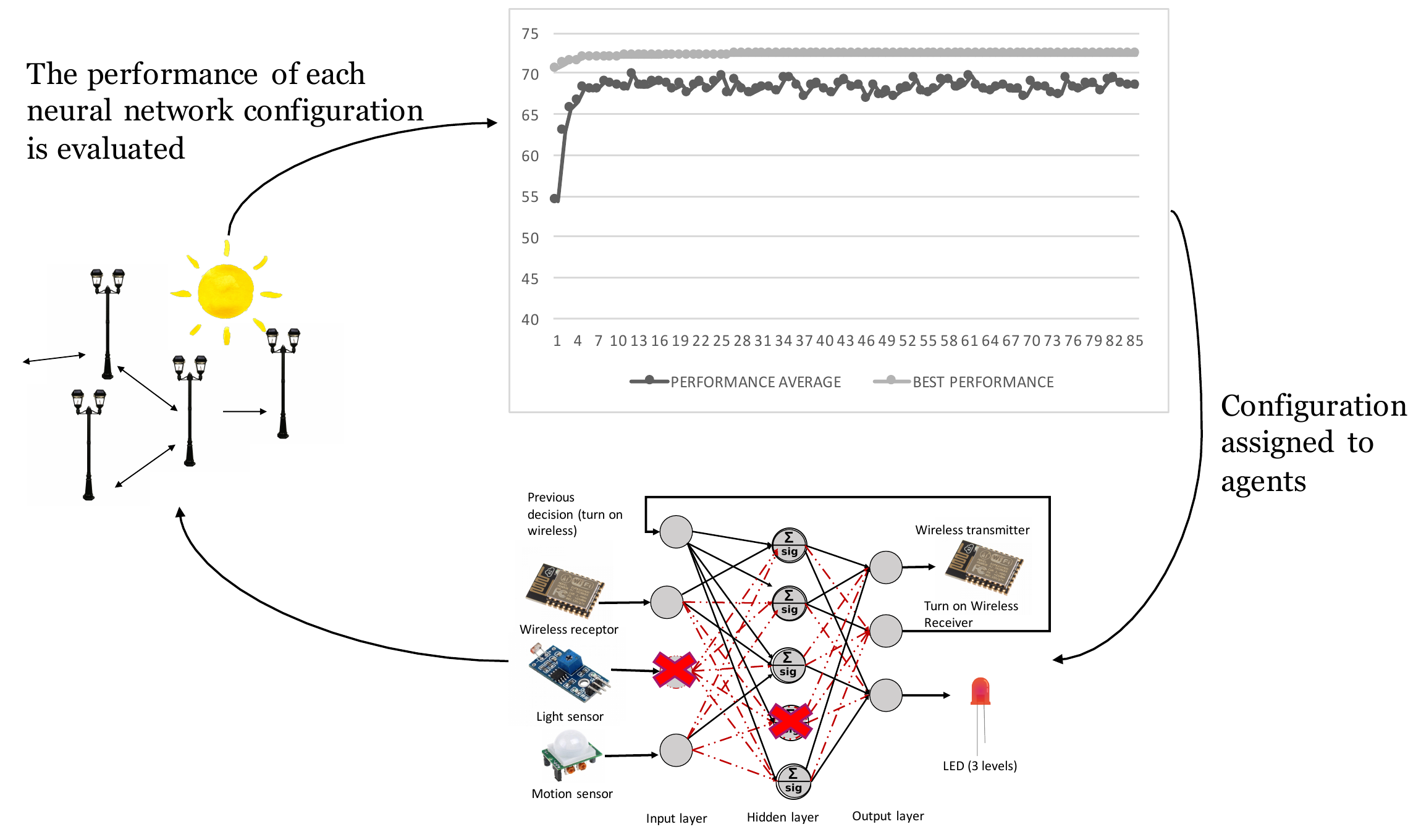}
	\caption{Learning the environment.}
	\label{figure:firstresult}
\end{figure}

As illustrated in Figure \ref{figure:firstresult}, in this first step, the scenario was bright during the entire period that the agents were learning about the environment. After some learning interactions, the agents developed an appropriate behavior to achieve their tasks in this version of the application. As a result, the neuroevolution-based algorithm discarded all the neuronal connections between the light sensor and the hidden units.

As the environment was always sunny, an obvious behavior for the agent is to turn off the light, whether a person was present or not. As expected, the agents produced this behavior. We were also expecting that the number of hidden neurons would be considerably decreased because of the simplicity of the task. But only one hidden neuron and one feature input were removed. This behavior occurred because the IoT expert selected a sigmoid function as an activation function.

As known, the output of the sigmoid function is not zero when its input is zero (i.e. if the sigmoid input is zero, the LED will be turned on). Thus, the learning algorithm found a configuration to assure that the LED remains turned off.

\subsection{Reconfiguring the set of features}
\subsubsection{The learning algorithm did not find an appropriate solution}
The IoT expert was expecting a fitness performance higher than 75\% and a simpler architecture to use in the real devices. But after several interactions of the learning algorithm with the environment, the highest fitness performance achieved by the learning algorithm was 72\%.

 As the human-in-the-loop was not satisfied with this result, he/she reconfigured the first set of features that was used to compose the IoT agents. For instance, an alternative choice of the activation function of the neural network was selected: namely, the binary activation function with threshold. As a result, the performance result quickly increased by more than 5\% and the human-in-the-loop obtained a simpler architecture.

\subsubsection{The environment unexpectedly changed}

\begin{figure}[!htb]
	\centering
	\includegraphics[width=7.9cm]{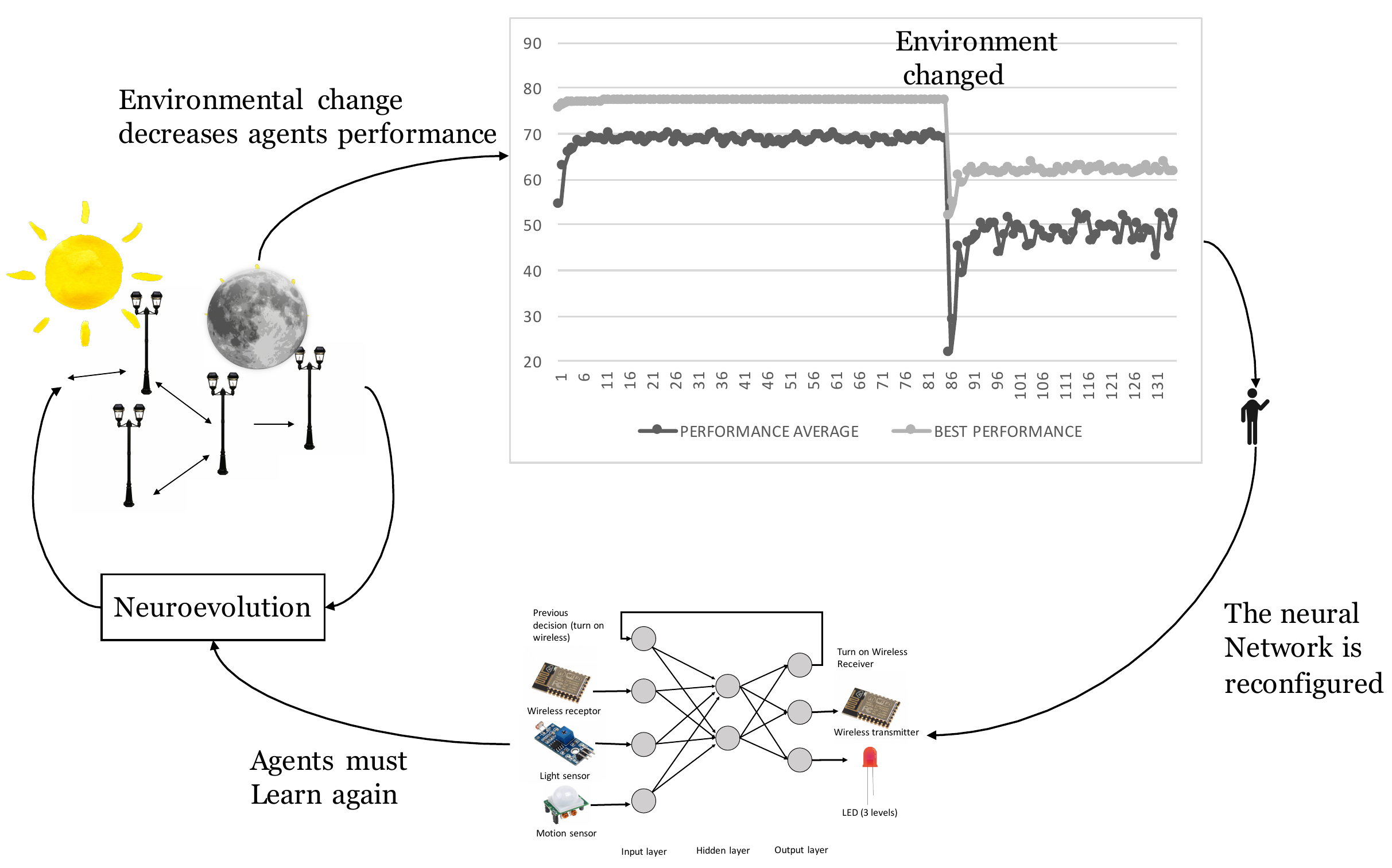}
	\caption{Reconfiguring the set of features.}
	\label{figure:secondlearninginteraction}
\end{figure}

After a time, a change in the environment occurred. Now, these agents are operating in an environment in which sometimes the background light can be bright and at other times dark. As a result, the performance of the set of agents considerably decreased, as shown in Figure \ref{figure:secondlearninginteraction}. The learning algorithm continued its training from its last state, but the current analysis architecture was not a viable option for this new situation.

The human-in-the-loop evaluated this decreased performance, and then reconfigured the system. For instance, the expert could have selected a new sensor, but he/she maintained the number of sensor inputs, but selected different variants for the neural network, such as ``two" as the maximum number of neurons in the hidden layer and the sigmoid activation function. Then, the learning algorithm was re-executed and the agents learned to cope with this environmental change.

\section{Related Work}

\cite{whiteson2005automatic,tan2009automated,diuk2009adaptive
	,nguyen2013online,ure2014distributed,del2017piecewise}  are some of the examples that apply feature selection to handle variability in learning agent-based systems. For example, \cite{diuk2009adaptive} propose an approach that uses reinforcement learning algorithms for structure discovery and feature selection while actively exploring an unknown environment. To exemplify the use of the proposed algorithm, the authors present the problem of a unique robot that has to decide which of multiple sensory inputs such as camera readings of surface color and texture, and IR sensor reading are relevant for capturing an environment’s transition dynamics.

However, most of these approaches do not address the problem of environmental- or user-based reconfiguration, in which a new set of features may be selected based on environmental changes, expanding or changing the search space of the group of learning agents. In addition, as most of these approaches load all features into the agent, they do not address the problem of dealing with mandatory, optional and alternative features.

\cite{sharifloo2016learning} provides one of the few solutions that propose an approach for feature selection and feature set reconfiguration. They presented a theoretical approach that proposes the use of reinforcement learning for feature selection, and a reconfiguration guided by system evolution where the user creates new features to deal with changes in  system requirements. They do not consider the changes that can happen dynamically in the environment, which can be handled by an automatic module by testing alternative choices of features.

In addition, most of these approaches do not characterize variability in their application domain. In fact, \cite{galster2014variability} observed that most approaches for variability handling are not oriented to specific domains. These approaches are potentially widely applicable. However, \cite{galster2014variability} consider that for a variability approach to cover complex domains, it is necessary to create domain-specific solutions. Therefore, \cite{galster2014variability} consider the extension of variability approaches for specific domains as a promising direction for future work.

\section{Contributions and Ongoing Work}
We provided an approach through which a software engineer with expertise in IoT agents co-worked with a  neuroevolutionary-based algorithm that can discard features. First, the software engineer provided the initial configuration of the agent-based system, using personal expertise to select a set of features. Then, a neuroevolutionary-based algorithm was executed to remove those features that were selected by the developer, but shown to be irrelevant to the application during the simulation.

However, after an unexpected environmental change that was not considered by the software engineer during the initial design time, the previous solution found by the neuroevolutionary algorithm stopped to work. Then, after evaluating the environmental changes, the software engineer had three options: i) to add a new feature to the feature model; ii) to select alternative choices of some features, including a different neural network architecture and properties, then starting the learning process again; and iii) to maintain the set of features and just reactivate the learning algorithm to continue from its last state.

In addition, to handling variability in learning for IoT agents, we identified the main variation points of these kinds of applications, including the variants that can be involved in a neural network design. We also provided a feature-oriented variability model, which is an established software engineering module.

The proposed approach is an example of a human-in-the-loop approach in which a machine-learning automated-procedure assists a software developer in completing his/her task. Our next step is to enable the use of a learning technique to reconfigure the set of features based on environmental changes automatically. As we proposed a hybrid architecture, we can use this learning technique only to reconfigure the variants related to one of the variation points, such as the neural network properties. In such an instance, we can have a human-in-the-loop responsible for handling the body and behavior variability of the IoT agents.


\section*{Acknowledgments}
 This work has been supported by CAPES scholarship/Program 194/Process: 88881.134630/2016-01 and the Laboratory of Software Engineering (LES) at PUC-Rio. It has been developed in cooperation with the University of Waterloo, Canada. Our thanks to CNPq, CAPES, FAPERJ and PUC-Rio for their support through scholarships and fellowships.

\appendix

\bibliographystyle{named}
\bibliography{sigproc}

\end{document}